\documentclass[onecolumn]{aastex631}

\usepackage{mathtools}
\usepackage{footnote}
\shorttitle{NH$_2$CHO formation by photolysis of CO:NH$_3$ ice mixtures}
\shortauthors{Chuang et al.}
\graphicspath{{./}{figures/}}

\begin{document}

\title{Formation of the simplest amide in molecular clouds: formamide (NH$_{2}$CHO) and its derivatives in H$_2$O-rich and CO-rich interstellar ice analogs upon VUV irradiation}

\author[0000-0001-6877-5046]{K.-J. Chuang}\altaffiliation{Current address: Laboratory for Astrophysics, Leiden Observatory, Leiden University, P.O. Box 9513, NL-2300 RA Leiden, the Netherlands.}
\affiliation{Laboratory Astrophysics Group of the Max Planck Institute for Astronomy at the Friedrich Schiller University Jena, Institute of Solid State Physics, Helmholtzweg 3, D-07743 Jena, Germany}
\email{chuang@strw.leidenuniv.nl}

\author[0000-0001-9393-0183]{C. Jäger}
\affiliation{Laboratory Astrophysics Group of the Max Planck Institute for Astronomy at the Friedrich Schiller University Jena, Institute of Solid State Physics, Helmholtzweg 3, D-07743 Jena, Germany}

\author[0000-0002-9816-3187]{S.A. Krasnokutski}
\affiliation{Laboratory Astrophysics Group of the Max Planck Institute for Astronomy at the Friedrich Schiller University Jena, Institute of Solid State Physics, Helmholtzweg 3, D-07743 Jena, Germany}

\author[0000-0002-7260-9742]{D. Fulvio}\altaffiliation{Current address: Istituto Nazionale di Astrofisica, Osservatorio Astronomico di Capodimonte, Salita Moiariello 16, I-80131 Naples, Italy.}
\affiliation{Max Planck Institute for Astronomy, Königstuhl 17, D-69117 Heidelberg, Germany}

\author[0000-0002-1493-300X]{Th. Henning}
\affiliation{Max Planck Institute for Astronomy, Königstuhl 17, D-69117 Heidelberg, Germany}



\begin{abstract}

The astronomical detection of formamide (NH$_2$CHO) toward various star-forming regions and in cometary material implies that the simplest amide might have an early origin in dark molecular clouds at low temperatures. Laboratory studies have proven the efficient NH$_2$CHO formation in interstellar CO:NH$_3$ ice analogs upon energetic processing. However, it is still under debate, whether the proposed radical–radical recombination reactions forming complex organic molecules remain valid in an abundant H$_2$O environment.
The aim of this work was to investigate the formation of NH$_2$CHO in H$_2$O- and CO-rich ices under conditions prevailing in molecular clouds. Therefore, different ice mixtures composed of H$_2$O:CO:NH$_3$ (10:5:1), CO:NH$_3$ (4:1), and CO:NH$_3$ (0.6:1) were exposed to vacuum ultraviolet photons in an ultra-high vacuum chamber at 10 K. Fourier-transform infrared spectroscopy was utilized to monitor in situ the initial and newly formed species as a function of photon fluence. The infrared spectral identifications are complementarily secured by a temperature-programmed desorption experiment combined with a quadrupole mass spectrometer. 
The energetic processing of CO:NH$_3$ ice mixtures mainly leads to the NH$_2$CHO formation, along with its chemical derivatives such as isocyanic acid (HNCO) and cyanate ion (OCN$^-$). The formation kinetics of NH$_2$CHO shows an explicit dependency on ice ratios and compositions; the highest yield is found in H$_2$O-rich ice. The astronomical relevance of the resulting reaction network is discussed.

\end{abstract}



\section{Introduction} \label{INTRODUCTION}
Complex organic molecules (COMs) have been widely observed in massive molecular clouds, high- and low-mass protostars, and cometary material (see reviews by \citealt{Herbst2009, Herbst2017, Altwegg2019}). Among the successfully identified O-, N-, and S-bearing COMs, some of them attract exceptional attention in the field of astrochemistry due to their potential role in the extraterrestrial origin of building blocks of life \citep{MunozCaro2002, Meierhenrich2005, Nuevo2008, Fulvio2021, Ioppolo2021}. 
For example, 
formamide (NH$_2$CHO) has been suggested to be an important precursor for the formation of nucleobases (e.g., adenine, uracil, thymine, cytosine, and guanine), which are found in RNA and/or DNA structures \citep{Saladino2006, Barks2010, Rotelli2016}.

In astronomical observations, NH$_2$CHO was first detected by \cite{Rubin1971} toward Sgr B2. Later, it has been observed by follow-up searches in various high-mass protostellar objects \citep{Bisschop2007, Halfen2011, Isokoski2013, Jones2013, Neill2014, Suzuki2018, Colzi2021}. 
NH$_2$CHO has also been identified toward several solar-mass stars 
 and shock regions 
\citep{Takahiro2012, Kahane2013, Mendoza2014, Lopez-Sepulcre2015, Taquet2015, Coutens2016, Imai2016, Lee2017, Oya2017, Marcelino2018}. A detailed list of astronomical observations can be found in the review by \cite{Lopez2019}. Although the detection of NH$_2$CHO in prestellar cores is still lacking, the common detection of the simplest amide in cometary comae, which have been considered to represent the pristine interstellar ice composition, suggests that it could have an early and cold origin from dense molecular clouds \citep{Bockelee1997, Biver2014, Goesmann2015, Altwegg2017, Drozdovskaya2019}.

Both gas-phase and solid-state formation routes of NH$_2$CHO have been discussed to explain its widespread and abundant detection in the aforementioned star-forming regions. The gas-phase reaction NH$_2$ + H$_2$CO $\rightarrow$ NH$_2$CHO + H has been theoretically suggested to be efficient under interstellar cloud conditions \citep{Barone2015, Skouteris2017}. However, its reaction rate constant is still under debate \citep{Song2016}. 
In addition, the gas-phase reaction CO + NH$_2$ $\rightarrow$ NH$_2$CO and the following, so-called “radical disproportionation” (i.e., 2NH$_2$CO $\rightarrow$ NH$_2$CHO + HNCO) have been reported in early high-temperature studies (150-400 ℃) simulating the primordial Earth's conditions \citep{Boden1970, Yokota1973, Hubbard1975}. 

\begin{table*}[t]\label{Table01}
	\caption{Overview of the Experiments Performed}             
	\centering                          
	\begin{tabular}{lcccclc}
		\hline
		\hline
		Exp & \textit{T} & \textit{N}(H$_2$O) & \textit{N}(CO) & \textit{N}(NH$_3$) &  Ratio & VUV Fluence \\
        & (K) & (molecules cm$^{-2}$) & (molecules cm$^{-2}$) & (molecules cm$^{-2}$) & H$_2$O:CO:NH$_3$ & (photons cm$^{-2}$) \\
        \hline
        CO:NH$_3$+VUV & 10 & - & 2.3$\times$10$^{16}$ & 3.6$\times$10$^{16}$ & \ 0 \ :\ 0.6:\ 1 & 8.2$\times$10$^{17}$ \\
        CO:NH$_3$+VUV & 10 & - & 1.4$\times$10$^{17}$ & 3.6$\times$10$^{16}$ & \ 0 \ : \ 4\ :\ 1 & 8.2$\times$10$^{17}$ \\
        H$_2$O:CO:NH$_3$+VUV & 10 & 2.6$\times$10$^{17}$ & 1.5$\times$10$^{17}$ & 2.9$\times$10$^{16}$ & 10 : \ 5 : 1 & 8.2$\times$10$^{17}$ \\
        H$_2$O:CO+VUV & 10 & 2.0$\times$10$^{17}$ & 1.5$\times$10$^{17}$ & - & \: 1 : \ 1 : 0 \ & 8.2$\times$10$^{17}$ \\
        H$_2$O:NH$_3$+VUV & 10 & 3.6$\times$10$^{17}$ & - & 3.3$\times$10$^{16}$ & 10 : \, 0 :\ 1 & 8.2$\times$10$^{17}$ \\
        \hline
    \end{tabular}
\end{table*}

In the solid phase, several low-temperature formation mechanisms of NH$_2$CHO have been investigated, including atomic addition, radical association with interstellar molecules (e.g., CO, NH$_3$, and H$_2$O), and radical-radical recombination reactions. 
It has been concluded that the successive hydrogenation of isocyanic acid (HNCO) is unlikely to form NH$_2$CHO on dust grains under interstellar conditions, because of a relatively low rate constant \citep{Song2016}. Furthermore, the dual-cyclic mechanism of H-atom abstraction and addition reactions connecting HNCO and NH$_2$CHO favors the hydrogen-poor species \citep{Haupa2019}. Following the gas-phase work reported in \cite{Hubbard1975}, the radical-molecule association of NH$_2$ + CO resulting in NH$_2$CO radicals has been 
adapted to explain the NH$_2$CHO formation in the energetic processing of CO:NH$_3$ ice mixture 
\citep{Hudson2000b, Bredehoft2017}. However, this pathway is questioned because of the absence of NH$_2$CO \citep{Jones2011}. Another radical-molecule association route comprising H$_2$CO and NH$_2$O, which is formed in situ through H-atoms addition/abstraction of NO/NH$_2$OH, has been proposed in \cite{Dulieu2019}, but further validation of its efficiency is still
desired. As an alternative, barrierless radical-radical recombination 
of NH$_2$ and HCO with proper geometric orientation has been suggested to form NH$_2$CHO in the solid phase \citep{Hudson2000b}. This formation mechanism has also been investigated under different energetic sources, such as fast ions, electrons, and UV as well as X-ray photons \citep{Jones2011, Abplanalp2016, Fedoseev2016, Ciaravella2019, Martin2020}. 

The solid-state chemistry of the CO:NH$_3$ ice mixture triggered by energetic processing has been extensively investigated in the past decades \citep{Milligan1965,  Ferris1974, Hagen1979, Grim1989, Demyk1998, Hudson2000b}.  Although NH$_2$CHO is a prevalent product detected upon UV photon or ion impacts, the main focii of these studies were more on the spectroscopy of [NH$_4]^+$[OCN]$^-$ and on the identification of refractory material of biological interests 
after a relatively long irradiation time. In the last ten years, quantitative studies of the product evolution as a function of energy have become available. For example, \cite{Jones2011} studied the energetic electron bombardment of the CO:NH$_3$ ice mixture at $\sim$12 K and reported the temporal profiles of HCO and NH$_2$CHO to support the suggested reaction pathway, i.e., HCO + NH$_2$ $\rightarrow$ NH$_2$CHO. More recently, a systematic study showed the dependence of the NH$_2$CHO formation on the CO:NH$_3$ ratios and temperatures (10-24 K) upon vacuum UV (VUV) irradiation \citep{Martin2020}. The derived kinetics of the NH$_2$CHO formation was further 
compared to the formation of the analog product CH$_3$CHO following similar recombination reactions (CH$_3$ + CHO) in the CH$_4$:NH$_3$ ice mixture and to the data obtained by \cite{Jones2011}. It has been concluded that both VUV photons and energetic electrons have very similar effects on the formation cross-section of formamide. In spite of these numerous studies, the validity of the proposed solid-state chemistry leading to complex species in an astronomically realistic icy environment comprising abundant H$_2$O ice is still under debate. More experimental studies are needed in order to understand the role of water in the COM formation, as mentioned in \cite{Lopez2019}. The concern is linked to a higher diffusion barrier of radicals and the limited orientations at the surface and in the bulk ice of H$_2$O \citep{Enrique2019, Lopez2019}. \cite{Agarwal1985} studied the UV photolysis of interstellar ice containing H$_2$O, CO, and NH$_3$ using a relatively high photon fluence. However, they found a list of large prebiotic compounds, with the notable exception of formamide. More recently, \cite{Ciaravella2019} reported the NH$_2$CHO formation upon X-ray irradiation of an ice mixture with a similar quantity of H$_2$O, CO, and NH$_3$ (1:0.9:0.7) at 13 K. 
Besides this, the potential role of abundant H$_2$O on the formation kinetics of formamide still remains unclear.

The previous studies mentioned above have devoted great effort into exploring each specific topic, such as OCN$^-$ formation/identification, NH$_2$CHO formation mechanisms, and kinetics of the newly formed N-bearing products. These topics are important to comprehensively understand the interstellar ice evolution in molecular clouds. However, they have not been experimentally investigated in a single chemical system considering the possible influence of the most abundant H$_2$O ice. The current laboratory study, for the first time, aims at verifying the formation of NH$_2$CHO and its chemical derivatives (e.g., HNCO and OCN$^-$) in CO:NH$_3$ ice mixtures with or without H$_2$O triggered by the cosmic ray induced secondary VUV photons (mainly H$_2$ emissions; \citealt{Gredel1989, Cruz-Diaz2014a}). The goal of this study is to reveal a potential chemical network involving the three chemically important species H$_2$O, CO, and NH$_3$ in interstellar ice and remove the doubts concerning the COM formation in H$_2$O-rich ice mantles. 
The present work is motivated by the importance of NH$_2$CHO in astrochemistry and highly requested quantitative data regarding the NH$_2$CHO formation in relevant ice compositions. In this study, the kinetic evolution of all newly formed products, including both N- and O-bearing species, is monitored in situ as a function of photon fluence. The experimental details are described in the next section. The results are presented in section 3 and discussed in section 4. The astronomical relevance of the experimental findings and conclusions are summarized in section 5.

\section{Experimental} \label{sec:Experiment}

\begin{table*}[]\label{Table02}
	\caption{Summary of IR peak Positions and Band Strength Values Used in This Work.}             
	\label{Table1}      
	\centering                          
    \begin{tabular}{ccccc}
    \hline
    \hline
    Species & Chemical Formula & IR Peak Position & Band Strength & References \\
    & & (cm$^{-1}$) & (cm molecule$^{-1}$) & \\
    \hline
    Carbon monoxide & CO & 2137 & 1.1$\times$10$^{-17}$ & \cite{Jiang1975, Bouilloud2015} \\
    Ammonia & NH$_3$ & 1040 & 2.0$\times$10$^{-17}$ & \cite{Hudson2022} \\
    Formamide & NH$_2$CHO & 1698 & 6.5$\times$10$^{-17}$ & \cite{Brucato2006} \\
    Isocyanic acid & HNCO & 2261 & 7.2$\times$10$^{-17}$ & \cite{vanBroekhuizen2004} \\
    Cyanate & OCN$^-$ & 2150 & 1.3$\times$10$^{-16}$ & \cite{vanBroekhuizen2004} \\
    Formaldehyde & H$_2$CO & 1724 & 9.6$\times$10$^{-18}$ & \cite{Schutte1993} \\
    Formyl radical & HCO & 1855 & 1.8$\times$10$^{-17}$ & \cite{Ryazantsev2017} \\
    Carbon dioxide & CO$_2$ & 2343 & 1.2$\times$10$^{-16}$ & \cite{Gerakines2015} \\
        \hline
    \end{tabular}
\end{table*}

All experiments have been performed in the Laboratory Astrophysics and Cluster Physics Group in Jena using the ultra-high vacuum (UHV) apparatus. The experimental details have been described in previous work \citep{Potapov2019}, and here only relevant information as well as further modifications are mentioned. The base pressure in the UHV chamber is $\sim5\times10^{-10}$ mbar at room temperature. A copper sample holder is mounted on the tip of a closed-cycle helium cryostat (ColdEdge CS-204) and positioned at the center of the UHV chamber. The temperature at the sample holder monitored by a silicon diode with $\leq$0.5 K accuracy can be varied between 10 and 330 K, and is controlled by a resistive heater manipulated by a Lakeshore temperature controller. The thermal contact between the cryostat and the substrate is improved using an indium foil to minutesimize the temperature difference. In this work, a KBr window, which has a high rate of transmission in the mid-IR range, has been used as a bare substrate that allows the intended ice samples to condense and the bulk ice chemistry triggered by VUV photons at 10 K to be studied. 

The VUV photons have been generated by a commercial deuterium lamp (Hamamatsu: H2D2-L11798), a closed-cycle (MgF$_2$-sealed) light source, and further guided to the ice sample through a MgF$_2$ UHV viewport installed on the main chamber. The transmission of a MgF2 window decreases from more than $\sim$88\% at 160.0 nm to $\sim$66\% at 121.6 nm (for 2 mm thickness). After calibrating the default emission spectrum provided by the Hamamatsu with the standard transmission of the MgF$_2$ window, the contribution of Ly$\alpha$ emission in the range of 115-128 nm drops significantly, 
making it negligible compared to the D$_2$ emissions at 128-170 nm. 
The spectral energy distribution with a stronger emission band at $\sim$160 nm represents the H$_2$ Lyman band system induced by cosmic rays interacting with hydrogen molecules in dense molecular clouds \citep{Prasad1983, France2005}. The UV photon chemistry induced by the H$_2$ Werner band system remains for future studies. Although the in situ measurement of UV flux at the sample position is still lacking, a lower photon flux is expected. The estimated UV flux ($\phi$) of $\sim$7.6$\times$10$^{13}$ photons cm$^{-2}$ s$^{-1}$ has been determined by comparing the formation rate constant ($k'$) of formamide in this work to the averaged formation cross-section ($\sigma$) value reported in \cite{Martin2020}, who used the same D$_2$ lamp under very similar experimental settings: $\phi$=$k'$/$\sigma$.

Pure CO (Westfalen, 99.97\%) or CO mixed with H$_2$O vapor (Sigma-Aldrich, UHPLC grade) and NH$_3$ (Air Liquide, 99.999\%) were simultaneously introduced into the main chamber through two separate all-metal leak valves and deposited onto the precooled KBr substrate, which is positioned at an angle of 30$^{\circ}$ to the deposition tubes and 90$^{\circ}$ to the IR beam. In order to investigate CO-poor, CO-rich, and H$_2$O-rich interstellar ices, three  different ice compositions were intentionally selected: (1) CO:NH$_3$=0.6:1, (2) CO:NH$_3$=4:1, and (3) H$_2$O:CO:NH$_3$=10:5:1. The deposited ice mixtures were then irradiated with UV photons at 10 K. The applied ice compositions and column densities are summarized in Table \ref{Table01}.

The growth of the deposited ices and the chemistry induced by the photolysis were monitored in situ using Fourier transform infrared spectroscopy (FTIR) in transmission mode in a range of 7500-400 cm$^{-1}$ with 1 cm$^{-1}$ resolution. The quantitative analysis of the IR spectra followed the commonly used method described in previous studies (see \citealt{Jones2011} and \citealt{Martin2020}) on the formamide formation, and is only briefly explained below. The selected IR absorption peaks were corrected by straight baseline subtraction, and the overlapping peaks were deconvoluted using Gaussian functions, which are both available in the \textit{OriginPro} software. Their corresponding absorption areas were obtained directly by Gaussian curve fitting. The absolute abundances of parent molecules and products were derived by converting IR absorption peak areas to column densities utilizing the modified Beer-Lambert Law \citep{Bennett2004, TerwisschavanScheltinga2018}, 
\begin{equation}\label{Eq0}
\textit{N} = \frac {\ln 10 \cdot \int \textit{Abs}(\nu) d(\nu)}{\textit{A'}}
\end{equation}
where \textit{N} is the column density in molecules cm$^{-2}$, \textit{Abs}($\nu$) is the IR absorbance, and \textit{A’} is the corresponding IR band strength in cm molecule$^{-1}$. The error bars were defined as the standard deviation of the Gaussian fit. This estimation does not account for uncertainties originating from the baseline subtraction procedure and IR absorption band strengths, which have been reported in the literature. The used IR band strengths are summarized in Table \ref{Table02}. The derived column densities of these species can be further refined as soon as more precise values become available.

After VUV irradiation of predeposited (H$_2$O:)CO:NH$_3$ ice mixtures for 180 minutes, the photon-processed ice samples were sublimated by a temperature-programmed desorption (TPD) experiment with a ramping rate of 5 K minute$^{-1}$. The desorbed species were analyzed with a quadrupole mass spectrometer (QMS; XT300M, Extorr Inc.). Characteristic desorption temperatures and fragmentation patterns induced by the ionization with 70 eV electrons were recorded in order to identify the newly formed products by comparing them with standard samples in the NIST database and literature. Such a QMS-TPD diagnostic technique has demonstrated a much higher sensitivity with respect to FTIR.

\section{Results} \label{RESULTS}
\subsection{NH$_{2}$CHO formation in CO:NH$_{3}$ ice mixtures}

\begin{figure}[t]
	\begin{center}
		\includegraphics[width=85mm]{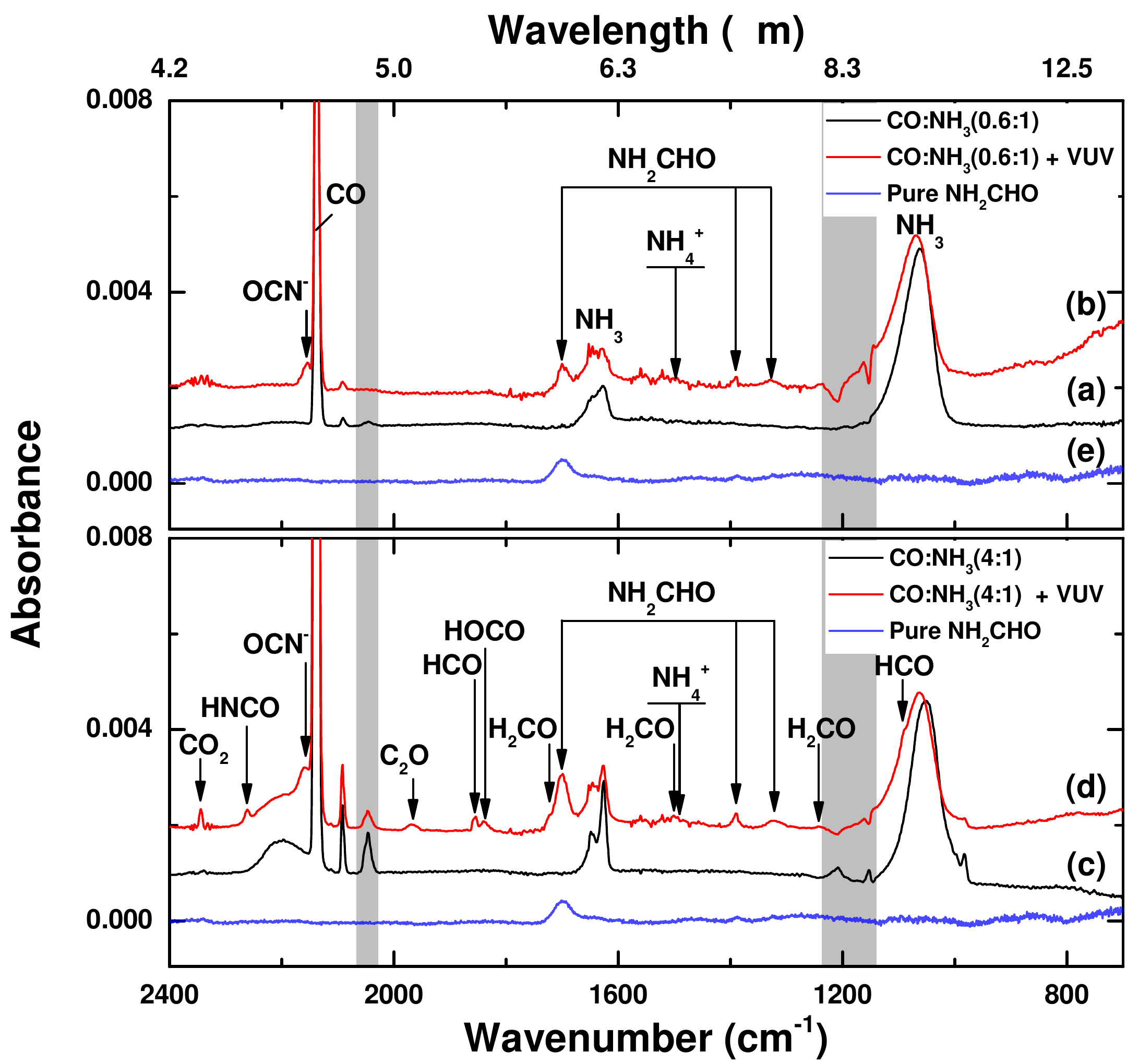}
		\caption{Top panel: IR spectra obtained after deposition of ice mixtures (a) CO:NH$_{3}$ (0.6:1) and (b) CO:NH$_{3}$ (0.6:1) after VUV irradiation for a fluence of 8.2$\times$10$^{17}$ photons cm$^{-2}$ (180 minutes) at 10 K. Bottom panel: IR spectra obtained after deposition of (c) CO:NH$_{3}$ (4:1) and VUV irradiation of (d) CO:NH$_{3}$ (4:1) ice mixture for a fluence of 8.2$\times$10$^{17}$ photons cm$^{-2}$ (180 minutes). IR spectra are offset for clarity. IR spectrum (e) is shown for pure NH$_2$CHO ice at 10 K. The shadowed areas are representative for the contamination with Ni(CO)$_{4}$ from the gas regulator and the FTIR defect features at 2046 and $\sim$1200 cm$^{-1}$, respectively \citep{meunier2019}.}
		\label{Fig1}
	\end{center}
\end{figure}
The top panel of Fig. \ref{Fig1} presents the IR absorption spectra of a CO:NH$_3$ (0.6:1) ice mixture obtained before (a) and after (b) VUV irradiation for 180 minutes at 10 K, as well as that of (e) pure NH$_2$CHO as a reference. Besides the absorption features corresponding to the parent molecules, such as CO at 2143 cm$^{-1}$ and NH$_{3}$ at 1030 cm$^{-1}$ as well as 1626/1649 cm$^{-1}$ (monomer/aggregate), several IR peaks were only observed upon photolysis of the deposited ice mixture in the spectrum (b), such as the absorption peaks at 1698, 1390, and 1324 cm$^{-1}$ corresponding to C=O stretching ($\nu_4$), C-H bending ($\nu_6$), and C-N stretching ($\nu_7$) modes of NH$_{2}$CHO, respectively \citep{King1971, Torrie1994, Brucato2006, Sivaraman2013}. Other less intense features of NH$_2$CHO are overlapped with IR peaks of NH$_3$. The absorption feature at 2155 cm$^{-1}$ is assigned to OCN$^{-}$ \citep{Demyk1998, Bernstein2000, Hudson2000b, Raunier2003, vanBroekhuizen2004, Gerakines2004}. Historically, it was labeled as "XCN" in astronomical observations and then successfully identified as the ionic species OCN$^{-}$ by comparing with laboratory (isotope-labeled) studies \citep{Grim1987, Schutte1997}. The counterpart cation NH$_4^+$ is shown at $\sim$1495 cm$^{-1}$. 

In the bottom panel of Fig. \ref{Fig1}, the IR absorption spectra of the CO:NH$_3$ (4:1) ice mixture obtained before (c) and after (d) VUV irradiation for 180 minutes at 10 K as well as of (e) pure NH$_2$CHO are shown. In spectrum (c), the full width at half maximum (FWHM) of the NH$_{3}$ symmetric deformation ($\nu_{2}$) mode becomes more narrow (10$\%$) compared to that in the CO-poor case (spectrum (a)), due to a cage effect of overabundant CO. This is also supported by the observed splitting of the antisymmetric and symmetric deformation modes ($\nu_{4}$) of the NH$_{3}$ monomer. All spectral characteristics clearly suggest that NH$_{3}$ resides in the CO-rich ice. 

Besides the previously labeled products, several new IR peaks are additionally present in spectrum (d). The IR feature at 2260 cm$^{-1}$ is assigned to HNCO ($\nu_2$), a precursor of OCN$^{-}$ \citep{Milligan1965, Teles1989, Lowenthal2002, vanBroekhuizen2004}. The possible isomer cyanate acid (HOCN) is absent, which is consistent with the previous studies suggesting a less efficient isomerization from HNCO to HOCN \citep{Theule2011,Jimenez2014}. The vibrational peaks at 1855 and 1090 cm$^{-1}$ originate from the CO stretching ($\nu_3$) and CH bending ($\nu_2$) modes of HCO, which is the first intermediate product of the CO hydrogenation scheme \citep{Ewing1960, Milligan1964}. The detection of HCO is consistent with the well-known H-atom addition reactions to CO ice, i.e., CO$\rightarrow$H$_{2}$CO$\rightarrow$CH$_{3}$OH \citep{Watanabe2002a, Fuchs2009}. The strongest C=O ($\nu_2$) vibrational band of H$_{2}$CO is clearly observable at 1722 cm$^{-1}$ along with other relatively weak absorptions at 1500 ($\nu_3$) and $\sim$1245 ($\nu_5$) cm$^{-1}$ \citep{Khoshkhoo1973, Nelander1980}. The most intense absorption peak of the saturated product CH$_{3}$OH (i.e., C-O stretching $\nu_8$) cannot be deconvoluted easily from the broad feature of parent NH$_{3}$ around 1030 cm$^{-1}$. In addition to reactions with free H atoms originating from the NH$_3$ photolysis, CO is expected to be photoexcited (i.e., CO$\rightarrow$CO*) and followed by a reaction with another CO resulting in the formation of CO$_{2}$, which is observed at 2343 cm$^{-1}$. However, this mechanism is considered inefficient compared to the reaction of the hydrocarboxyl radical with atomic hydrogen HOCO + H $\rightarrow$ CO$_{2}$ + 2H \citep{watanabe2002b, Ioppolo2011, Chen2014}. Here, we cannot exclude the contribution of this reaction to the CO$_{2}$ formation, given the tiny peak of HOCO observed at 1840 cm$^{-1}$. 
\begin{figure}[]
	\begin{center}
		\includegraphics[width=85mm]{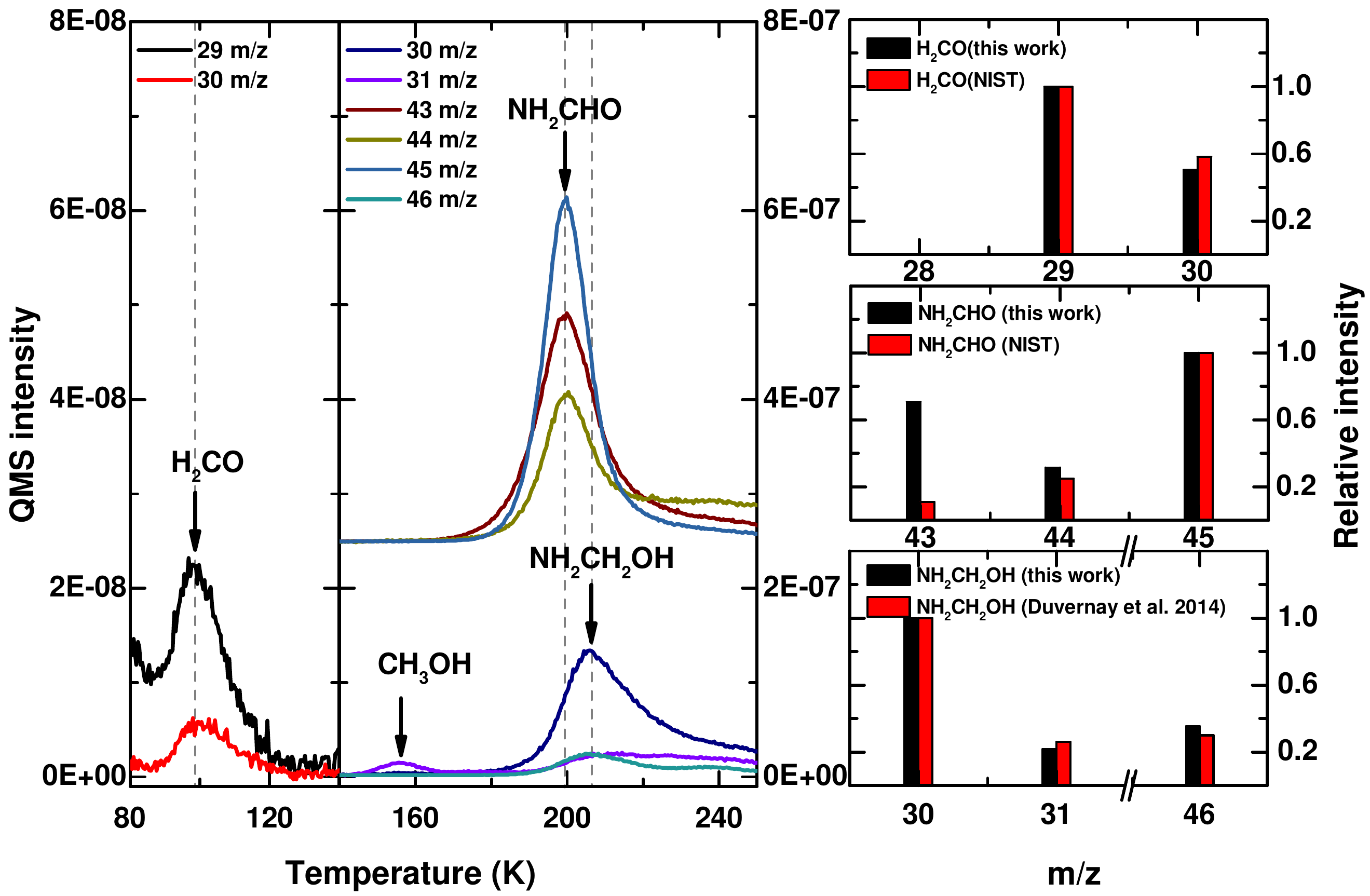}
		\caption{Left: Ion signals obtained during TPD experiment after VUV irradiation of CO+NH$_{3}$ (4:1) for a fluence of 8.2$\times$10$^{17}$ photons cm$^{-2}$ (180 minutes) at 10 K. Only relevant \textit{m/z} channels are shown, and they have been shifted for clarity. The dashed lines indicate the peak positions of the corresponding molecules. Right: Comparison of the electron-ionization fragmentation patterns for the QMS-TPD peaks at $\sim$100, $\sim$200, and $\sim$205 K (black) with standard data of H$_2$CO, NH$_2$CHO, and NH$_2$CH$_2$OH.}
		\label{Fig2}
	\end{center}
\end{figure}

The left panel of Fig. \ref{Fig2} presents the ion signals, which are obtained during the TPD experiment after VUV irradiation of CO:NH$_{3}$(4:1) ice at 10 K. Only ion signals at relevant masses are selected to complementarily secure the IR identification. For example, H$_{2}$CO is observed through the mass desorption peaks 29 and 30 \textit{m/z} at $\sim$100 K, corresponding to the most intense fragment and the molecular mass, respectively. The CH$_{3}$OH desorption is demonstrated by the most intense mass fragment of 31 \textit{m/z} at $\sim$156 K. The other two desorption peaks at higher temperatures are due to N-bearing species. The peak at $\sim$200 K (starting from $\sim$170 K; i.e., 43, 44, and 45 \textit{m/z}) and the peak at $\sim$205 K (starting from $\sim$180 K; i.e., 30, 31, and 46 \textit{m/z}) are consistent with the reported desorption temperature of NH$_{2}$CHO and NH$_{2}$CH$_{2}$OH, respectively \citep{Bossa2009, Ligterink2018, Martin2020}. A somehow different desorption profile has been observed for the mass signal of 43 \textit{m/z}, which starts increasing at a slightly lower temperature (i.e., $\sim$165 K) than those observed for the mass signals of 44 and 45 \textit{m/z}. This points to an additional contribution of 43 \textit{m/z} from HNCO, which is expected to desorb at $\sim$185 K \citep{Fischer2002, Fedoseev2015c}. Furthermore, the identifications are confirmed by comparing their (selected) mass fragmentation patterns induced by electron impact in the right panel of Fig \ref{Fig2}. The relative intensities between different mass signals are in good agreement with standard electron (70 eV) ionization patterns as available from the NIST database for H$_{2}$CO and NH$_{2}$CHO and from the study by \cite{Duvernay2014} for NH$_{2}$CH$_{2}$OH, except for the 43 \textit{m/z}, which has contributions from both HNCO and NH$_{2}$CHO. The experimental results from the ion signals during the TPD experiment support the IR-spectral assignments of H$_{2}$CO and NH$_{2}$CHO and further point out the possible formation of their saturated derivatives such as CH$_{3}$OH and NH$_{2}$CH$_{2}$OH, whose IR vibrational features might largely overlap with their precursors and whose intensities are below the FTIR detection limits.    

\begin{figure}[]
	\begin{center}
		\includegraphics[width=85mm]{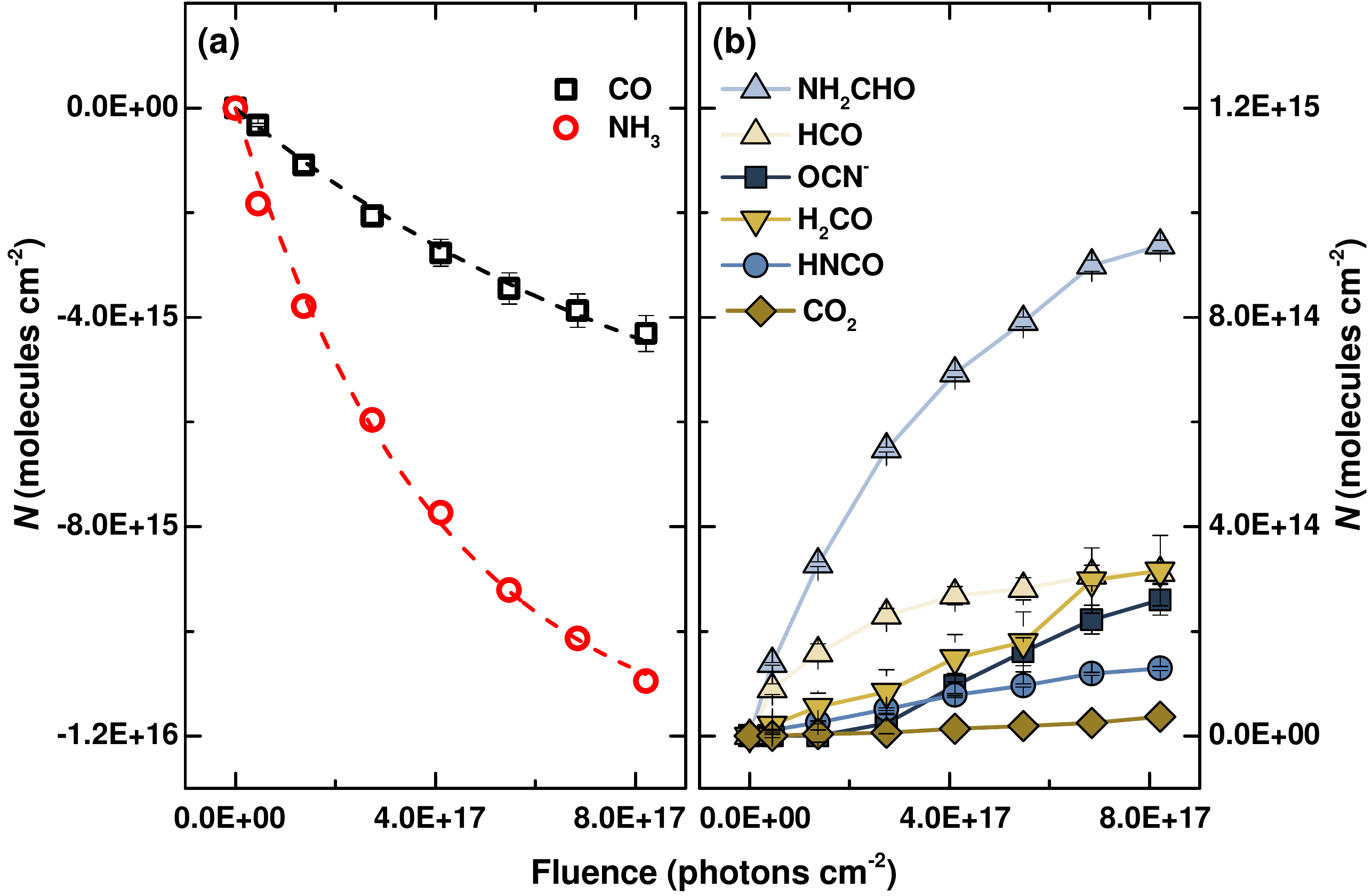}
		\caption{Evolution of (a) parent molecules and (b) newly formed products after VUV irradiation of CO:NH$_{3}$ (4:1) ice mixture over a fluence of 8.2$\times$10$^{17}$ photons cm$^{-2}$ (180 minutes). The dashed lines present the fitting results, and the solid lines connecting data are only for clarity.}
		\label{Fig3}
	\end{center}
\end{figure}

\subsection{Kinetic Evolution of CO:NH$_3$ Ice Mixture under VUV Irradiation}
In addition to the qualitative identifications of the photolysis products, the absolute column density (\textit{N}) of the parent molecules and newly formed species were monitored in situ as a function of photon fluence. In Fig. \ref{Fig3}, the evolutions of parent molecules (e.g., CO and NH$_{3}$ in Fig. \ref{Fig3}(a)) and major products (e.g., NH$_{2}$CHO, HCO, OCN$^{-}$, H$_{2}$CO, HNCO, and CO$_{2}$ in Fig. \ref{Fig3}(b)) after VUV irradiation of the CO:NH$_{3}$ (4:1) ice (fluence of 8.2$\times$10$^{17}$ photons cm$^{-2}$) are shown. The consumption of CO and NH$_{3}$ is fit by a single exponential equation:
\begin{equation}\label{Eq01}
\Delta\textit{N}\text{(molecules)}=\alpha(1-\text{exp}(\sigma\cdot \phi\cdot t))\\,
\end{equation} 
where $\alpha$ is the saturation value (i.e., the maximum change when reaching the equilibrium state) in molecules cm$^{-2}$, $\sigma$ is the photolysis cross-section in cm$^{2}$ photon$^{-1}$, $\phi$ is the UV flux in photons cm$^{-2}$ s$^{-1}$, and \textit{t} is the irradiation time in seconds. The derived cross-section values are (1.0±0.2)$\times$10$^{-18}$ and (2.5±0.2)$\times$10$^{-18}$ cm$^2$ photon$^{-1}$ for CO and NH$_{3}$, respectively. The value of NH$_3$ is larger than that of CO, due to the direct photodissociation of NH$_3$ forming NH$_2$ and H. 

In Fig. \ref{Fig3}(b), NH$_{2}$CHO and HCO are the two products that form most efficiently at the beginning of VUV irradiation. 
The abundances of NH$_{2}$CHO and HCO increase rapidly up to a fluence of 2.7$\times$10$^{17}$ photons cm$^{-2}$. The formation cross-section values of NH$_2$CHO and HCO are (2.9±0.2)$\times$10$^{-18}$ and (6.0±0.7)$\times$10$^{-18}$ cm$^{2}$ photon$^{-1}$, respectively. Once the HCO production rate slows down, a similar saturation behavior is also observed in NH$_{2}$CHO. The obtained ratio of NH$_{2}$CHO/HCO shows a relatively constant value of 
2.7±0.2 after reaching a fluence of 2.7$\times$10$^{17}$ photons cm$^{-2}$ implying a chemical correlation between these two. The formation of H$_{2}$CO and HNCO is also observed upon VUV processing, but with much lower production yields compared to NH$_{2}$CHO. Unlike the aforementioned species, the production of OCN$^{-}$ and CO$_{2}$ is considered as a high-order generation of photolysis products, given the different kinetic curves and the delay of formation. The abundances were only detectable after reaching a fluence of 2.7$\times$10$^{17}$ photons cm$^{-2}$. However, at the end of the photolysis experiment, the absolute column density of OCN$^{-}$ was higher than that of HNCO. Similar product evolution as a function of photon fluence was also found in the CO:NH$_3$ (0.6:1) ice mixture for NH$_{2}$CHO, H$_{2}$CO, and OCN$^{-}$, but their intermediate products HCO and HNCO were absent.  

\begin{figure}[b]
	\begin{center}
		\includegraphics[width=85mm]{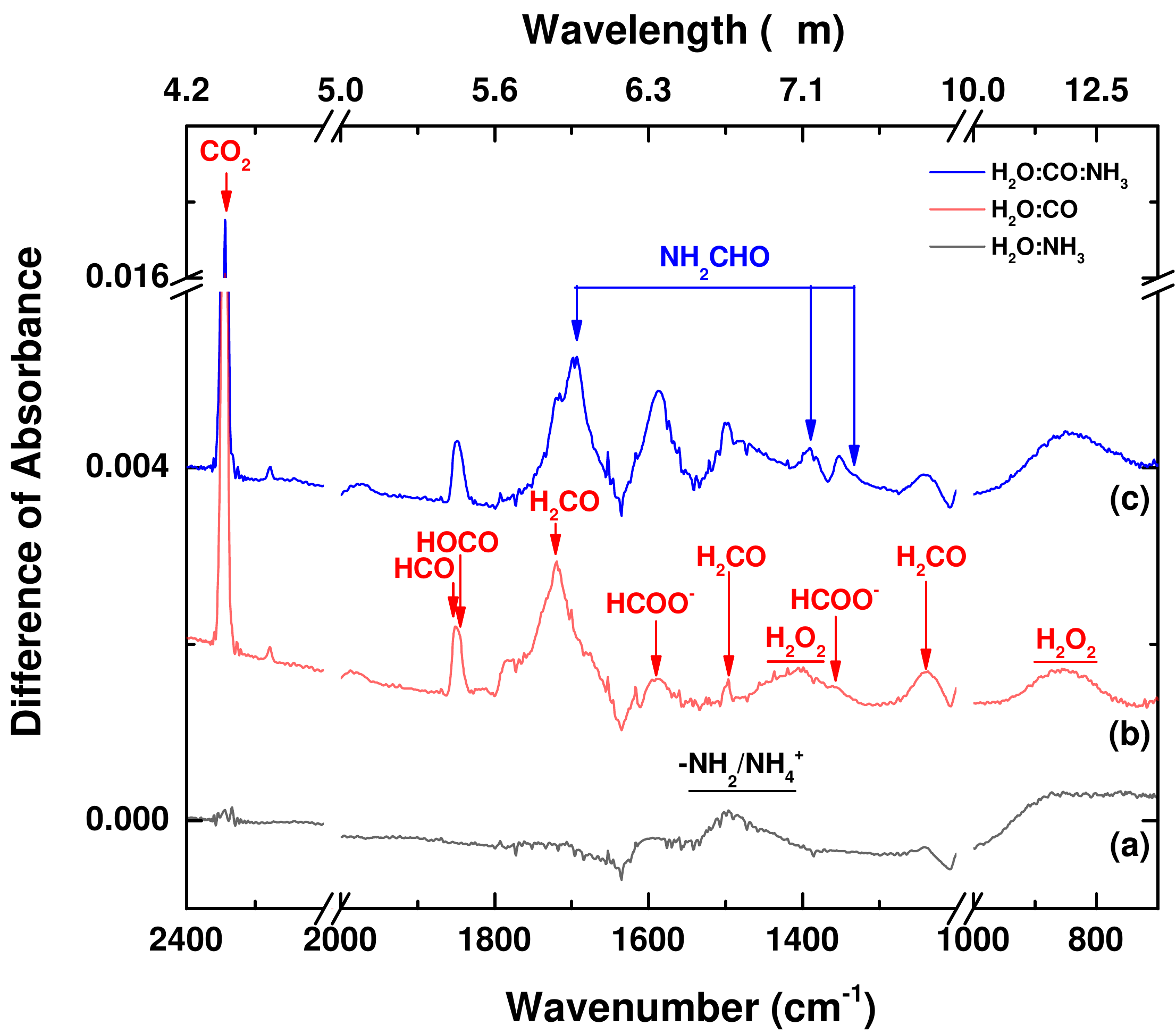}
		\caption{IR difference spectra obtained before and after UV irradiation of (a) H$_{2}$O:NH$_{3}$ (10:1), (b) H$_{2}$O:CO (1:1), and (c) H$_{2}$O:CO:NH$_{3}$ (10:5:1) ice mixtures for a fluence of 8.2$\times$10$^{17}$ photons cm$^{-2}$ at 10 K. IR spectra are offset for clarity.}
		\label{Fig4}
	\end{center}
\end{figure}

\subsection{NH$_{2}$CHO Formation in H$_2$O-rich Ice Mixture}

Figure \ref{Fig4} shows the difference IR absorption spectra before and after 180-min VUV irradiation of (a) H$_{2}$O:NH$_{3}$ (10:1), (b) H$_{2}$O:CO (1:1), and (c) H$_{2}$O:CO:NH$_{3}$ (10:5:1) ice mixtures at 10 K. The newly formed products are present in positive peaks and the parent molecules are present in negative features, which are intentionally omitted in Fig. \ref{Fig4}, except for H$_{2}$O at $\sim$1633 cm$^{-1}$. The control experiments (a) H$_{2}$O:NH$_{3}$ (10:1) and (b) H$_{2}$O:CO (1:1) provide spectral evidence to identify possible carriers of the newly formed peaks, which are exclusively detected in the H$_{2}$O:CO:NH$_{3}$ ice mixture (spectrum (c) in Fig. \ref{Fig4}). 

As shown in spectrum (a), VUV photolysis of H$_{2}$O:NH$_{3}$ ice mixture only results in a broad N-H bending feature around 1497 cm$^{-1}$, which has been assigned to NH$_{2}$ and/or NH$_{4}^{+}$ in [NH$_4]^{+}$[OH]$^{-}$ salt \citep{Schutte1999, Zheng2008}. The peaks of the possible product (NH$_2$OH) originating from OH + NH$_2$ are absent. This is probably due to a low formation yield or high destruction cross-section of NH$_2$OH upon VUV-photon impact \citep{Fedoseev2015b}. In spectrum (b), the new IR features are assigned to C-bearing species appearing upon VUV irradiation of H$_{2}$O:CO. Besides CO$_{2}$, H$_{2}$CO is another C-bearing product, presenting absorption bands at 1719, 1498, and 1244 cm$^{-1}$. The IR features of the intermediate radicals HCO and HOCO (i.e., the precursors of H$_{2}$CO and CO$_{2}$ , respectively) are largely overlapped, as shown at $\sim$1853/1846 cm$^{-1}$ \citep{Milligan1971, Bennett2010}. The spectral deconvolution of these two peaks was realized by Gaussian curve fitting and further confirmed during the TPD experiment because these two species have different desorption temperatures (IR absorptions at 1853 and 1846 cm$^{-1}$ disappeared at 15-30 and 30-45 K, respectively). The IR features of the formate anion (HCOO$^{-}$) are visible at 1587 and 1355 cm$^{-1}$ \citep{Milligan1971, Bennett2010}. This suggests a possible acid-base reaction between HCOOH and H$_2$O \citep{Schutte1999, Hudson2000b}. All aforementioned features shown in spectra (a) and (b) are also unambiguously present in spectrum (c) of the VUV irradiated H$_{2}$O:CO:NH$_{3}$ (10:5:1) ice mixture. In addition, the newly appeared IR features present in spectrum (c) must be attributed to photolysis products containing CN bonds; the IR peaks at 1693, 1392, and 1333 cm$^{-1}$ have been assigned to NH$_{2}$CHO. However, OCN$^{-}$ or HNCO, the chemical derivatives of NH$_{2}$CHO, are absent.

\begin{figure}[t]
	\begin{center}
		\includegraphics[width=85mm]{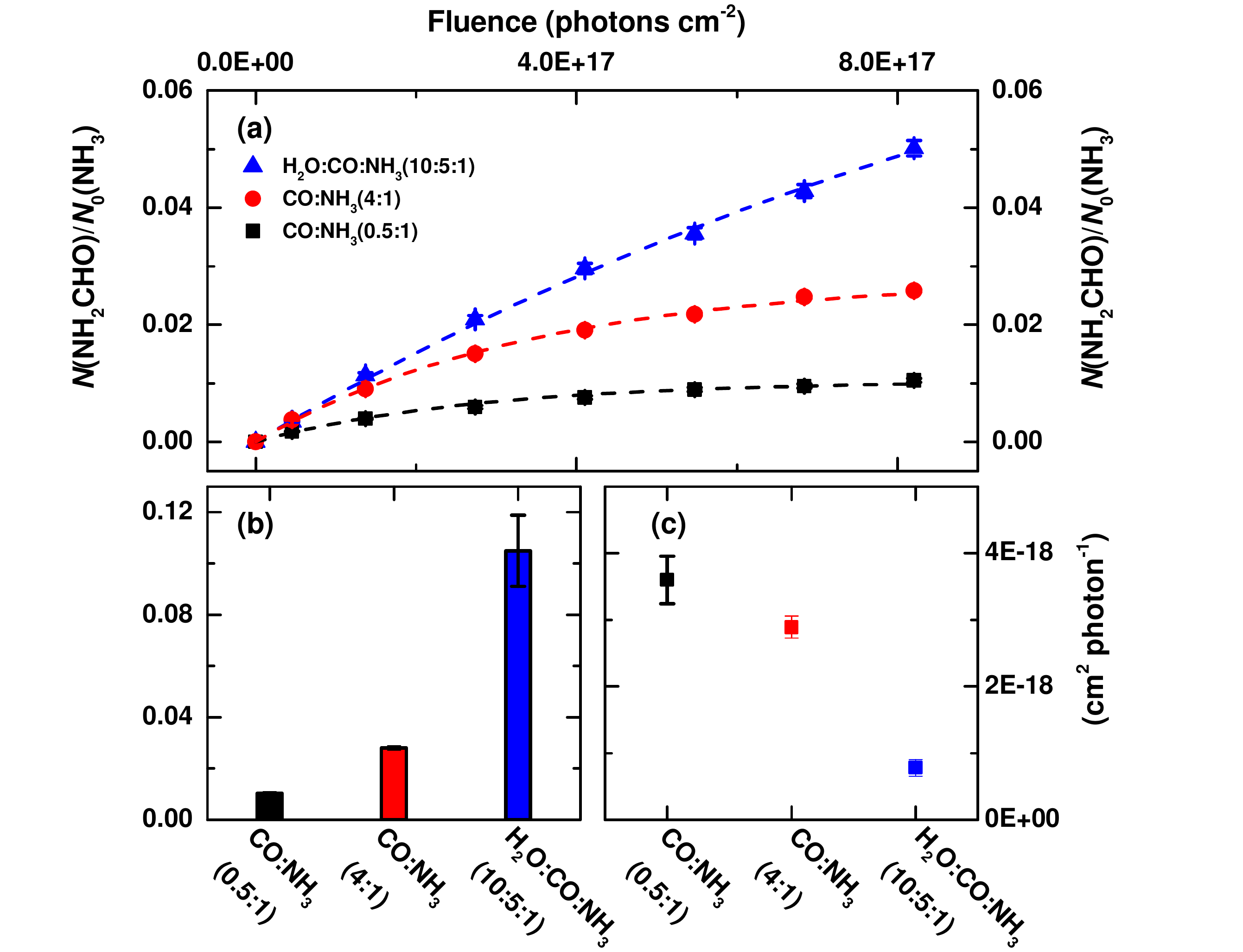}
		\caption{Upper panel: (a) Abundance evolution of the newly formed NH$_2$CHO in the VUV photolysis of three investigated ice mixtures: CO:NH$_{3}$ (0.6:1), CO:NH$_3$ (4:1), and H$_{2}$O:CO:NH$_{3}$ (10:5:1), over a fluence of 8.2$\times$10$^{17}$ photons cm$^{-2}$. 
		The dashed lines present the best fitting. Bottom panel: The derived fitting values of (b) saturation, $\alpha$, and (c) formation cross-section, $\sigma$, for NH$_{2}$CHO are present for three investigated CO:NH$_3$ ice mixtures}
		\label{Fig5}
	\end{center}
\end{figure}

The kinetic analysis shows that NH$_{2}$CHO is one of the first-generation products and is observed in all studied CO:NH$_{3}$ ice mixtures with or without the presence of H$_{2}$O. The evolution of \textit{N}(NH$_{2}$CHO) normalized by the initial parent species (i.e., \textit{N}$_0$(NH$_{3}$)) as a function of photon fluence and the fit according to Eq. \ref{Eq01} are shown in Fig. \ref{Fig5}(a). The corresponding values of the fitting, such as the saturation $\alpha$ (unitless) and the formation cross-section $\sigma$ (cm$^{2}$ photon$^{-1}$) are presented in Fig. \ref{Fig5}(b) and (c), respectively. It is important to note that the derived fitting outcomes are only considered as effective results that represent the combination of product's formation and destruction channels. The photon-induced cross-section values ($\sigma$) that indicate the efficiency of reaching the steady state are (3.6±0.4)$\times$10$^{-18}$, (2.9±0.2)$\times$10$^{-18}$, and (0.8±0.1)$\times$10$^{-18}$ cm$^{2}$ photon$^{-1}$ for CO:NH$_{3}$ (0.6:1), CO:NH$_{3}$ (4:1), and H$_{2}$O:CO:NH$_{3}$ (10:5:1) ice mixtures, respectively. Although a relatively low formation cross-section was obtained for the H$_2$O-rich experiment, the highest saturation value ($\alpha$) of 0.10±0.01 was found in the H$_2$O:CO:NH$_{3}$ (10:5:1) ice. In the CO:NH$_{3}$ (4:1) and CO:NH$_{3}$ (0.6:1) experiments, these values are considerably smaller, i.e., 0.028±0.001 and 0.001±0.001, respectively. The experimental results suggest that the NH$_{2}$CHO formation is strongly influenced by initial ice compositions and by the number of photons absorbed in the bulk ice. Ice mixtures, which are H$_{2}$O- or CO-rich, may play an important role in preventing the NH$_{2}$CHO destruction by absorbing incoming photons. Moreover, UV irradiation of H$_{2}$O ice can also lead to the formation of mobile (nonthermal) H atoms that can diffuse in the bulk ice and chemically convert photofragments, such as NHCHO or NH$_{2}$CO, back to NH$_{2}$CHO. The shielding from further destruction explains the absence of HNCO and OCN$^{-}$ in the H$_{2}$O:CO:NH$_{3}$ (10:5:1) experiment and enhances the NH$_2$CHO abundance as shown in Fig \ref{Fig5}(a). 

\begin{figure}[]
	\begin{center}
		\includegraphics[width=70mm]{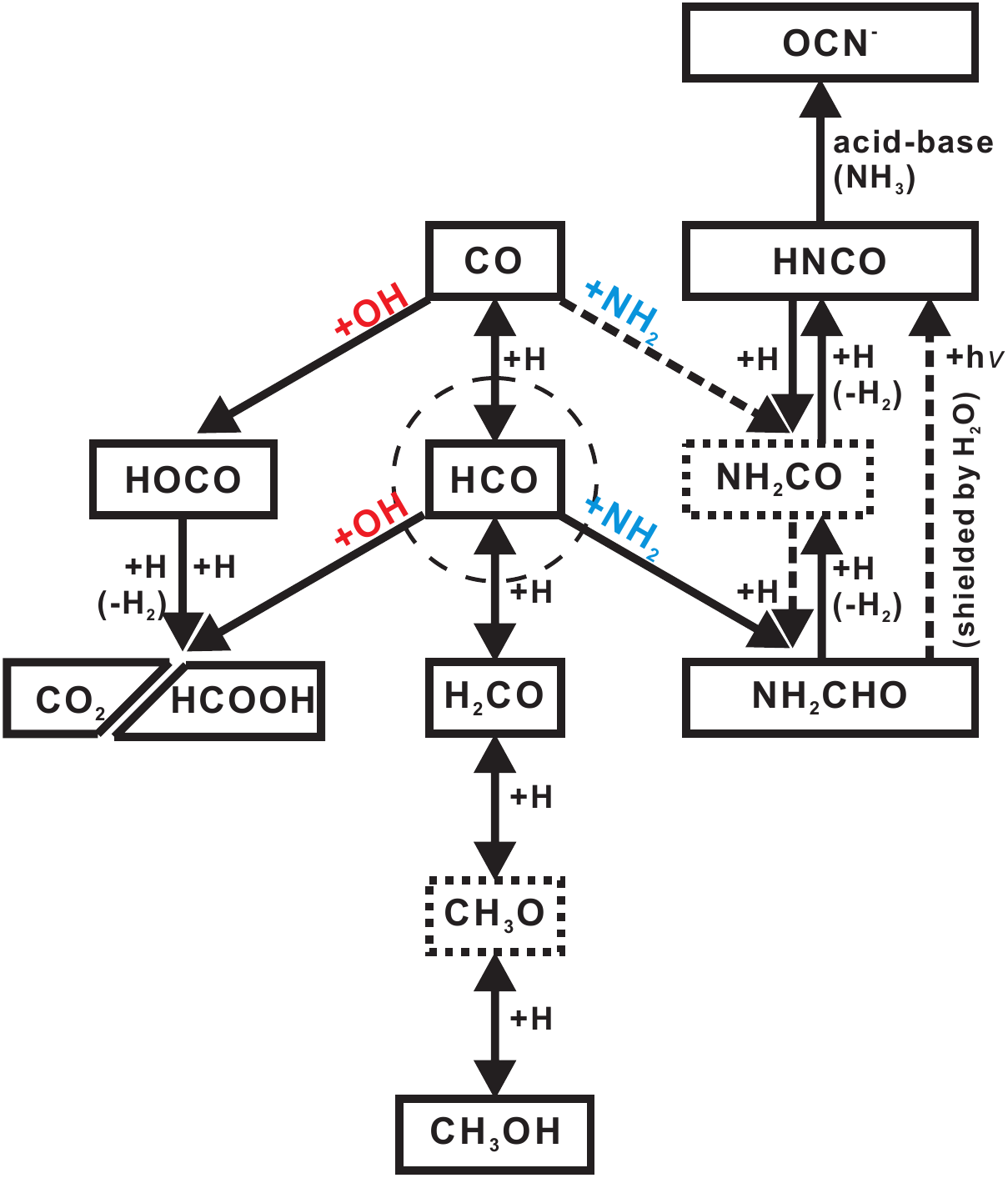}
		\caption{Proposed reaction diagram linking interstellar simple molecules and the newly formed N-bearing COMs formed at 10 K under VUV irradiation of H$_2$O:CO:NH$_3$ ice mixtures. The dashed boxes represent undetected intermediate species.}
		\label{Fig6}
	\end{center}
\end{figure}

\section{Discussion}
In this work, since there are two MgF$_2$ windows between the ice sample and the UV-photon source, the photon energy distribution is mainly characterized by $\sim$160 nm ($\sim$79\%) with a tiny contribution from Ly$\alpha$ ($\sim$6\%) as mentioned in the Experimental section. Therefore, the ionization of NH$_3$ by photons having an energy above 10 eV (i.e., $<$124 nm) is minutesimal compared to the photodissociation of NH$_3$. We only focus on the ice chemistry originating from reactions between neutral radicals in solid state in the following discussion. The possible ion/electron chemistry induced by photoionization in binary ice containing NH$_3$ is much more complex and has been experimentally investigated by \cite{Demyk1998} and \cite{Hudson2000b}. On the other side, the direct photodissociation of CO is forbidden due to the higher threshold energy required (i.e., 11.09 eV). However, CO has relatively strong absorption cross-section values (up to $\sim$1.5$\times10^{-17}$ cm$^2$ photon$^{-1}$) in the range of 127-157 nm coinciding with the molecular (D$_2$/H$_2$) emission ($\sim$130-165 nm). These photons can excite some CO molecules into the first electronically excited state (A$^1\Pi$ $\leftarrow$ X$^1\Sigma^+$; CO*), which carries excess energy of up to 7.9 eV \citep{Lu2005, Mason2006, Cruz-Diaz2014a}. In contrast, NH$_{3}$, with a much lower threshold energy of 4.4 eV, is expected to be efficiently dissociated upon the impact of VUV photons resulting in amino radicals (NH$_2$) and H atoms via the dissociation reaction \citep{Okabe1967}:
\begin{equation}
\label{Eq02}
\text{NH$_3$} \xrightarrow{\text{+h}\nu} \text{NH$_2$ + H}.
\end{equation}
Therefore, in a CO:NH$_3$ ice mixture, the consumption of CO is through either the so-called nonthermal desorption following the photon energy transfer from CO* to surface CO or chemical reactions with H atoms or NH$_2$ in the ice. For example, the interactions between CO and H atoms has been proposed to form HCO via H-atom addition reactions \citep{Watanabe2002a}: 
\begin{equation}
\label{Eq03}
\text{CO + H} \rightarrow \text{HCO}. 
\end{equation}
The newly formed HCO can immediately recombine with surrounding NH$_2$ radicals, resulting in NH$_2$CHO via the radical-radical recombination:
\begin{equation}
\label{Eq04}
\text{HCO + NH$_2$} \rightarrow \text{NH$_2$CHO}. 
\end{equation}
In addition, the successive hydrogenation of HCO could also take place, forming H$_2$CO:
\begin{equation}
\label{Eq05}
\text{HCO + H} \rightarrow \text{H$_2$CO}. 
\end{equation} 
The two HCO consumption channels (i.e., reactions (\ref{Eq04}) and (\ref{Eq05})) compete with each other. However, reaction (\ref{Eq04}) can occur when HCO and NH$_2$ radicals are nearby, and reaction (\ref{Eq05}) requires additional H atoms from another NH$_3$ photodissociation. Therefore, the formamide route is favored under the current experimental conditions and the derived ratio of \textit{N}(NH$_{2}$CHO)/\textit{N}(H$_{2}$CO) is $>$3 at steady state in all CO:NH$_{3}$ experiments. For each NH$_3$ dissociation event, one may expect that, after CO hydrogenation (reaction (\ref{Eq03})), NH$_{2}$ is the only radical near the newly formed HCO. The prompt recombination following reaction (\ref{Eq04}) is expected without further diffusion. The second H atom requires the photodissociation of another NH$_{3}$ and additional diffusion. It is important to note that radical-radical recombination reactions could also lead back to NH$_3$ and CO through the direct H transfer \citep{Enrique2019}. However, the repetitive NH$_3$ dissociation followed by the CO hydrogenation effectively favors the formation of formamide.   

Besides interactions between radicals and ground-state molecules, a fraction of electronically excited carbon monoxide (CO*) can directly react with NH$_2$ radicals followed by H-atom recombination, providing an additional formation pathway to NH$_2$CHO through the reactions:
\begin{equation}
\label{Eq06}
\text{CO* + NH$_2$} \rightarrow \text{NH$_2$CO}. 
\end{equation}
and
\begin{equation}
\label{Eq07}
\text{NH$_2$CO + H} \rightarrow \text{NH$_2$CHO}. 
\end{equation}
A higher activation barrier reported for reaction (\ref{Eq06}) with respect to that of reaction (\ref{Eq03}) (i.e., 15 kJ mol$^{-1}$ versus 8 kJ mol$^{-1}$, in the gas phase) can be easily compensated by the absorbed VUV-photon energy. However, if NH$_2$CO is formed as an intermediate product, it is strongly expected to be converted backward, forming HNCO \citep{Noble2015, Haupa2019}.

The chemistry of electronically excited CO with other strongly bound molecules, such as H$_2$ and N$_2$ with high activation barriers has been experimentally investigated in the solid state by \cite{Chuang2018b} and \cite{Martin2020}. The interactions between CO* and NH$_3$ might also be relevant in the present study. However, the presence of available CO* in the ice mixture strongly depends on the impacting photon fluence and the time-scale of energy dissipation in the bulk ice. Additionally, the direct proton transfer from NH$_3$ to CO could also contribute to the formation of NH$_2$CHO, given that the reaction barrier of 314.5 kJ mol$^{-1}$ calculated in current work for the isolated CO-NH$_3$ complex at the b3lyp/6-311G+(d,p) level can be easily overcome by the absorption of photons. Despite the nondetection of the vibrational feature of the CO-NH$_3$ complex ($\sim$2145 cm$^{-1}$; \citealt{Lundell1998}), this formation route cannot be excluded. 

The continued UV irradiation leads to further chemical modification of first-generation products. For example, the photolysis of NH$_{2}$CHO results in the HNCO formation through the H$_2$ loss channel \citep{Lundell1998, Duvernay2005}:
\begin{equation}
\label{Eq08}
\text{NH$_2$CHO} \xrightarrow{\text{+h}\nu} \text{HNCO + H$_2$}. 
\end{equation}
A very similar photodissociation of H$_2$CO is also expected, enriching the content of reactive radicals like HCO, H, or even CO. Moreover, the dehydrogenation pathway through H-atom abstraction reactions has also been reported via \citep{Haupa2019,Suhasaria2020}
\begin{equation}
\label{Eq09}
\text{NH$_2$CHO} \xrightarrow{\text{+H(-H$_2$)}} \text{NH$_2$CO} \xrightarrow{\text{+H(-H$_2$)}} \text{HNCO}. 
\end{equation}

The product HNCO can react with the surrounding NH$_3$ ice and convert into OCN$^{-}$, through the acid-base reaction:
\begin{equation}
\label{Eq010}
\text{HNCO + NH$_3$} \leftrightarrow \text{[NH$_4]^+$[OCN$]^-$}. 
\end{equation}
This is supported by observing a delay of the OCN$^{-}$ formation shown in Fig. \ref{Fig3}. The efficient conversion from the neutral isocyanic acid (HNCO) to the cyanate ion (OCN$^-$) has been experimentally confirmed at 10 K \citep{Jimenez2014}.

These secondary products (e.g., HNCO and OCN$^-$) are absent in the photolysis of the H$_2$O-rich ice mixture (H$_2$O:CO:NH$_3$=10:5:1). Instead, the abundance of NH$_2$CHO rapidly increases with photon fluence. NH$_2$CHO, which is formed by either reactions (\ref{Eq03})-(\ref{Eq04}) or (\ref{Eq06})-(\ref{Eq07}), is largely preserved in H$_2$O ice instead of being photodissociated into simpler products. 

In addition to the formation of dehydrogenated species, H$_2$CO could also be hydrogenated into saturated molecules through successive addition reactions of H atoms:
\begin{equation}
\label{Eq11b}
\text{H$_2$CO} \xrightarrow{\text{+2H}} \text{CH$_3$OH}. 
\end{equation}
However, hydrogenation of NH$_2$CHO forming NH$_2$CH$_2$OH has been shown to be less likely \citep{Noble2015}. Since aminomethanol is only tentatively observed in the QMS-TPD data, its formation could originate from the thermal chemistry \citep{Duvernay2014}:

\begin{equation}
\label{Eq11a}
\text{NH$_3$} + \text{H$_2$CO} \rightarrow \text{NH$_2$CH$_2$OH}. \end{equation}


As a consequence of the H$_2$O dissociation, some OH radical association reactions might take place as well. The interaction between OH radicals and CO is expected to result in the intermediate product HOCO through a barrierless reaction:
\begin{equation}
\label{Eq12}
\text{OH + CO} \rightarrow \text{HOCO}. 
\end{equation}
The hydrogenation of HOCO leads to the formation of CO$_2$ and HCOOH \citep{Ioppolo2011, Qasim2019}:
\begin{subequations}
\begin{equation}
\label{Eq13a}
\text{HOCO + H} \rightarrow \text{CO$_2$ + H$_2$} 
\end{equation}
and
\begin{equation}
\label{Eq13b}
\text{HOCO + H} \rightarrow \text{HCOOH}. 
\end{equation}
\end{subequations}
Alternatively, HCOOH is also formed through the recombination of HCO with OH radicals:
\begin{equation}
\label{Eq14}
\text{HCO + OH} \rightarrow \text{HCOOH}. 
\end{equation}

A similar acid-base reaction has also been reported for a HCOOH ice mixed with H$_2$O- or NH$_3$-rich environments, leading to ionic species \citep{Schutte1999, Hudson2000b, Bergner2016}:
\begin{equation}
\label{Eq15}
\text{HCOOH + H$_2$O (or NH$_3$)} \rightarrow \text{[HCOO$]^-$[H$_3$O]$^+$ (or [NH$_4$]$^+$)}. 
\end{equation}

The proposed reactions induced by VUV irradiation of astronomically relevant ice mixtures containing H$_2$O, CO, and NH$_3$ are summarized in Fig. \ref{Fig6}. 

\section{Astrochemical Implications and Conclusions}

The present laboratory study demonstrates the solid-state chemistry of forming NH$_2$CHO and its derivatives, including HNCO and [OCN]$^-$[NH$_4]^+$, by the VUV photolysis of CO:NH$_3$ ice mixtures at 10 K. We investigated the proposed formation mechanism involving HCO radicals and NH$_2$ radicals in two different interstellar-relevant ice compositions in molecular clouds: H$_2$O-rich and CO-rich ice analogs. The parent molecules and newly formed products were monitored in situ by IR spectroscopy as a function of photon fluence and complementarily confirmed by mass spectrometry during the TPD experiments. Besides the identifications of the first-generation products in both ice scenarios, the further interactions between NH$_2$CHO and energetic H atoms triggered by continuous VUV irradiation result in hydrogen-poor HNCO (or OCN$^-$ when reacting with NH$_3$). The kinetic analysis of N-bearing products offers a hint at the solid-state chemical network from the most abundant species (H$_2$O, CO, CO$_2$, and NH$_3$) after energetic processing in molecular clouds before evaporation.

In the early stage of molecular clouds (1$\leq A_{V}\leq$5), the gas density increases to 10$^4$ cm$^{-3}$ leading to a decrease of the cloud temperature down to $\sim$20 K. Atomic reactions are considered to dominate the surface chemistry on submicron-sized interstellar dust grains. For example, in translucent clouds (\textit{A}$_{V}\geq$3), H atoms and O atoms diffuse on the surfaces of dust grains, forming H$_2$O ice via the reactions O$\xrightarrow{\text{+H}}$OH$\xrightarrow{\text{+H}}$H$_2$O. This occurs along with other H-atom addition reactions to N and C atoms, leading to NH$_3$ and CH$_4$, respectively \citep{Hidaka2011, Fedoseev2015b, Qasim2020}. In the meantime, a fraction of gaseous CO is also expected to participate in the first H$_2$O-rich ice chemistry, leading to CO$_2$ formation \citep{Ioppolo2011}. As the gas density continuously increases in dense clouds, the temperature of dust grains drops further(i.e., there is a negative correlation between \textit{A}$_{V}$ and temperature; \citealt{Hocuk2017}), allowing more CO to condense onto the previous H$_2$O-rich ice layer.  This is the so-called “CO catastrophically freeze-out stage,” forming a second CO-rich ice layer \citep{Boogert2015}. With the significantly increased lifetime of H atoms on surfaces, the H-atom addition reactions play an essential role in converting inorganic CO into (complex) organic species, such as aldehydes and alcohols \citep{Watanabe2002a, Fuchs2009, Chuang2016, Chuang2017,  Fedoseev2017, Simons2020}. The accumulation of interstellar ice is expected to take place within 10$^5$ yr, based on theoretical estimations (i.e., 10$^9$/\textit{n}$_{\text{(H)}}$, where \textit{n}$_{\text{(H)}}$=10$^4$ cm$^{-3}$; see \citealt{Willacy1998}). 
Besides the surface chemistry dominated by atom(radical)-molecule reactions, the cosmic rays (CR) 
and the secondary UV photons 
induced by CR interacting with H$_2$ have been proven to act as efficient chemical triggers to facilitate the molecular complexity in the ice mantles in the remaining 10$^{5}$-10$^{7}$ yr, as the typical lifetime of molecular cloud has been estimated up to $\sim$10$^7$ yr \citep{Chevance2020}. Considering higher penetration depths of UV or CRs than the thickness of ice mantles, in contrast to atoms, these energetic sources can efficiently trigger the bulk ice chemistry involving the preserved molecules in the deep ice mantles, both in H$_2$O-rich and CO-rich ice layers \citep{Cruz-Diaz2014a, Cruz-Diaz2014b}.

The investigated photochemistry of ices leading to N-bearing organic species is expected to take place in different astronomically relevant ice layers. The applied UV fluence is comparable to the accumulated UV photons in molecular clouds, i.e., 3$\times$10$^{17-18}$ photons cm$^{-2}$ assuming a UV flux of 10$^{3-4}$ photons cm$^{-2}$s$^{-1}$ and a cloud lifetime of 10$^{7}$ yr \citep{Prasad1983, Cecchi1992, Shen2004}. 
In the H$_2$O-rich ice mixture, the interactions between CO and H$_2$O photofragments (e.g., OH radicals and H atoms) mainly contributed to the formation of C-bearing species, providing additional chemical routes to enrich the abundance of CO$_2$, H$_2$CO, HCOOH/HCOO$^-$, HCO, and HOCO in H$_2$O ices besides (non-energetic) atom addition reactions. The only N-bearing species formed in the ice was NH$_2$CHO. HNCO (or OCN$^-$) was not observed, suggesting that the newly formed NH$_2$CHO was most likely protected in the H$_2$O-rich environment. The experimental findings demonstrate that the NH$_2$CHO formation pathway proposed for the photolysis of CO:NH$_3$ ice mixtures is still valid in interstellar H$_2$O-rich ice layers. Apparently, the binding/diffusing energy of radicals in H$_2$O-rich ice can be overcome in the UV-photon induced ice chemistry. Moreover, this reaction channel takes place simultaneously with other reactions, such as CO+2H $\rightarrow$ H$_2$CO and CO+OH $\rightarrow$ HOCO induced by VUV-photons. In the interstellar CO-rich ice layer, the UV-photon induced production of C-bearing species is limited due to the scarcity of H$_2$O photofragments. CO$_2$ and H$_2$CO are minor products. Their precursor species, such as HCO and HOCO, are trapped in CO ice. The N-bearing products include NH$_2$CHO, HNCO, and OCN$^-$. 

In astronomical observations, the vibrational feature of OCN$^-$ (4.64 $\mu$m) was first detected in the solid phase by \cite{Lacy1984} and later identified by \cite{Demyk1998}. The correlation between HNCO and OCN$^-$ has been established \citep{Hudson2000b, Jimenez2014}. The present laboratory study also supports the chemical link between these two species at low temperatures without thermal processing. The absence of HNCO in the interstellar H$_2$O-rich ice probably implies fast acid-base reactions involving a proton transfer, which may not be limited to NH$_3$. An attempt to search for the solid-state IR absorption features of NH$_2$CHO has been reported by \cite{Schutte1999} and \cite{Raunier2004}, but only resulted in a tentative identification. However, NH$_2$CHO and its chemical derivative HNCO have been detected simultaneously in cometary material, which is believed to inherit the most pristine ice species in molecular clouds \citep{Biver2014, Altwegg2017}. Future observation of N-bearing organic molecules, including NH$_2$CHO and its (de-)hydrogenated species (e.g., HNCO and NH$_2$CH$_2$OH) in interstellar ice mantles using the JWST is desired in order to reveal the full chemical fingerprints of N-bearing species in star-forming regions. Astronomical observations show a strong abundance correlation between HNCO and NH$_2$CHO in star-forming regions (see Figure 1 in \citealt{Lopez2019}). The chemical transformation from NH$_2$CHO to HNCO has been found in this work. However, the derived ratio of NH$_2$CHO/HNCO $>$ 1, which is in contradiction to observations, suggests that there are additional destruction routes, such as H-atom abstraction, high-energetic photons, and cosmic ray dissociation, influencing products' degree of hydrogenation. This work acts as a case study on the UV photolysis of interstellar ice analogs containing CO:NH$_3$ in an H$_2$O-rich environment. A systematic comparison of the conversion rate between NH$_2$CHO and HNCO induced by different (non-)energetic sources is highly desired to better understand the full interstellar ice evolution.

In conclusion, based on the experimental results on the photolysis chemistry of CO:NH$_3$ ice mixtures in H$_2$O-rich and CO-rich ice analogs, the main findings of this work can be summarized as below:
\begin{enumerate}
\item VUV irradiation of CO:NH$_3$ ice mixtures in H$_2$O-rich and CO-rich environments at 10 K shows an efficient formation pathway leading to N-bearing molecules (NH$_2$CHO, HNCO, and OCN$^-$) under molecular cloud conditions.
\item The formation kinetics of NH$_2$CHO in H$_2$O-rich and CO-rich ice are derived (Fig. \ref{Fig5}). It shows a strong dependency on the initial ice compositions. The highest production yield of NH$_2$CHO is observed in the H$_2$O-rich ice mixture.
\item NH$_2$CHO is reported as a first-generation product of CO:NH$_3$ ice mixtures upon VUV irradiation. The continuous photon impact or the photon-induced formation of free (energetic) H atoms from H$_2$O and NH$_3$ can further influence the final product composition including the formation of HNCO and OCN$^-$.    
\end{enumerate}


\begin{acknowledgements}
    Th.H. acknowledges support from the European Research Council under the Horizon 2020 Framework Program via the ERC Advanced Grant Origins 83 24 28. Authors thank the Leiden Ice Database for providing the solid-state IR spectrum of formamide.  
\end{acknowledgements}

\bibliography{Ref}{}
\bibliographystyle{aasjournal}



\end{document}